\documentclass[11pt, a4paper]{article}

\usepackage[margin=1in]{geometry}      
\usepackage[utf8]{inputenc}            
\usepackage{graphicx}                  
\usepackage{hyperref}                  
\usepackage{multirow}                  
\usepackage{rotating}                  
\usepackage{tikz}                      
\usetikzlibrary{positioning, shapes, arrows.meta}
\usepackage{booktabs}                  
\usepackage{natbib}                    

\title{The Generative AI Paradox: GenAI and the Erosion of Trust, the Corrosion of Information Verification, and the Demise of Truth}

\author{Emilio Ferrara$^{1,2,3,*}$}

\date{
    \footnotesize
    $^{1}$ Thomas Lord Department of Computer Science, University of Southern California (USC)\\
    $^{2}$ Annenberg School for Communication, University of Southern California (USC), Los Angeles, CA, USA\\
    $^{3}$ Information Sciences Institute (ISI), University of Southern California (USC), Marina del Rey, CA, USA\\[2ex]
    $^{*}$Correspondence: \href{mailto:emiliofe@usc.edu}{emiliofe@usc.edu}
}

\begin{document}

\maketitle

\begin{abstract}
Generative AI (GenAI) now produces text, images, audio, and video that can be perceptually convincing at scale and at negligible marginal cost. While public debate often frames the associated harms as ``deepfakes'' or incremental extensions of misinformation and fraud, this view misses a broader socio-technical shift: GenAI enables \emph{synthetic realities}; coherent, interactive, and potentially personalized information environments in which content, identity, and social interaction are jointly manufactured and mutually reinforcing. We argue that the most consequential risk is not merely the production of isolated synthetic artifacts, but the progressive erosion of shared epistemic ground and institutional verification practices as synthetic content, synthetic identity, and synthetic interaction become easy to generate and hard to audit. This paper (i) formalizes synthetic reality as a layered stack (content, identity, interaction, institutions), (ii) expands a taxonomy of GenAI harms spanning personal, economic, informational, and socio-technical risks, (iii) articulates the qualitative shifts introduced by GenAI (cost collapse, throughput, customization, micro-segmentation, provenance gaps, and trust erosion), and (iv) synthesizes recent risk realizations (2023--2025) into a compact case bank illustrating how these mechanisms manifest in fraud, elections, harassment, documentation, and supply-chain compromise. We then propose a mitigation stack that treats provenance infrastructure, platform governance, institutional workflow redesign, and public resilience as complementary rather than substitutable, and outline a research agenda focused on measuring \emph{epistemic security}. We conclude with the \emph{Generative AI Paradox}: as synthetic media becomes ubiquitous, societies may rationally discount digital evidence altogether, raising the cost of truth for everyday life and for democratic and economic institutions.
\end{abstract}

\vspace{1em}

\begin{figure}[t]
    \centering
    \includegraphics[height=.4\linewidth]{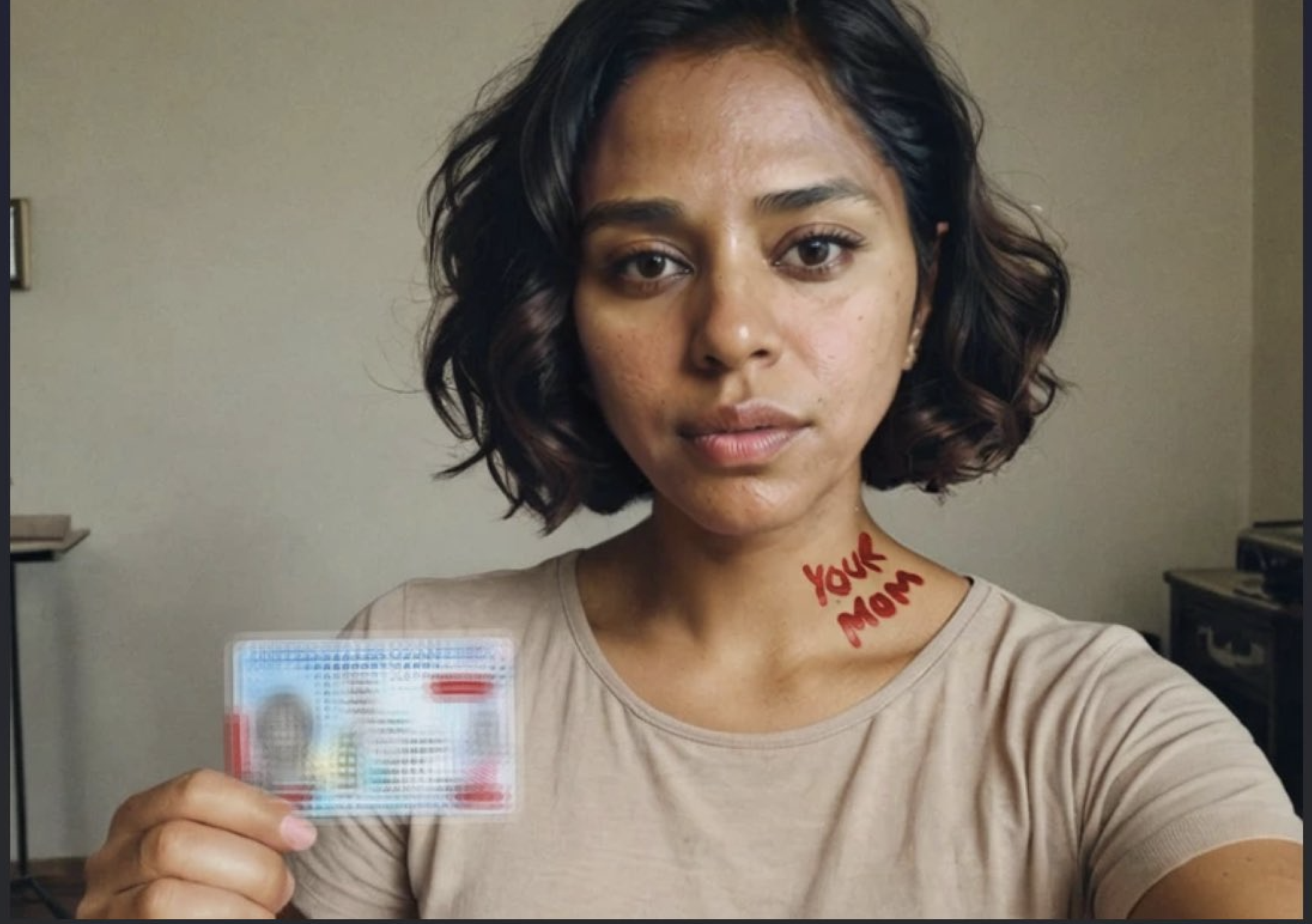}
    \includegraphics[height=.4\linewidth]{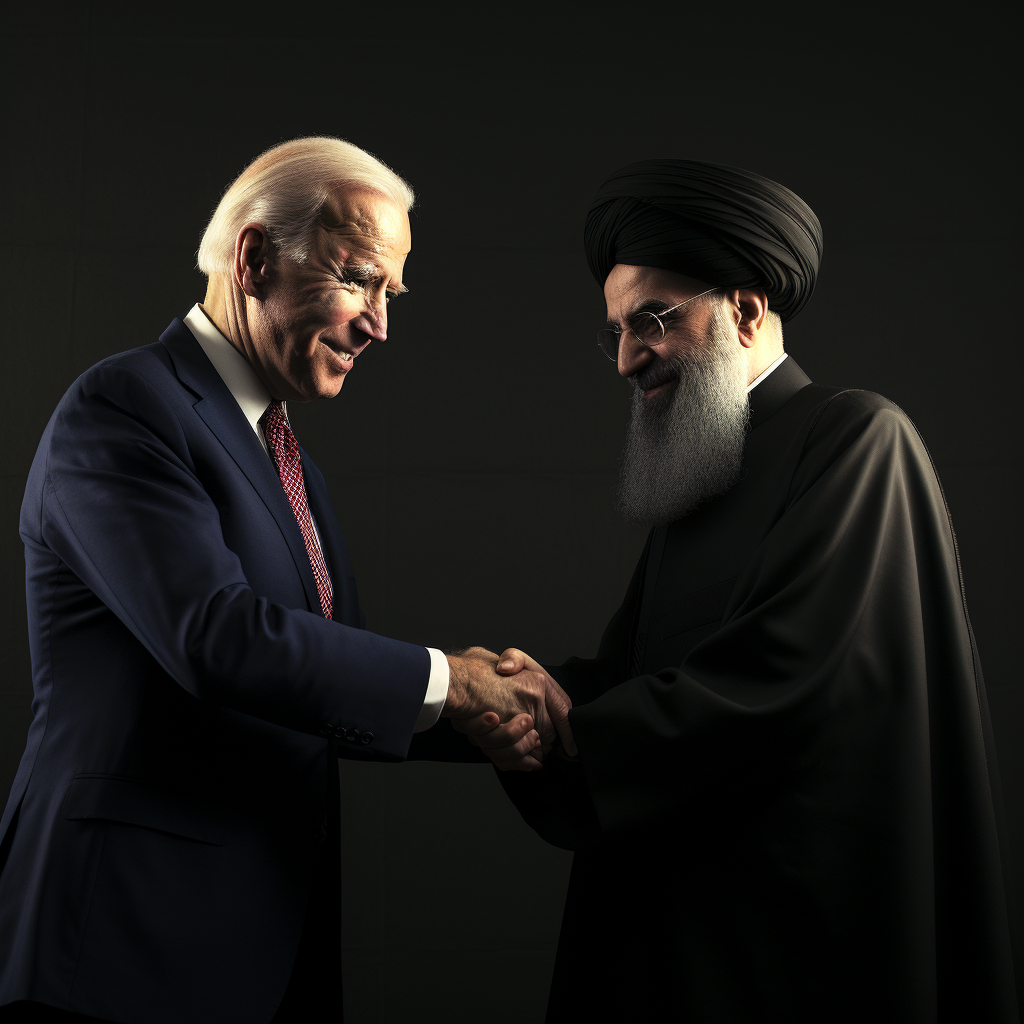}
    \includegraphics[height=.4\linewidth]{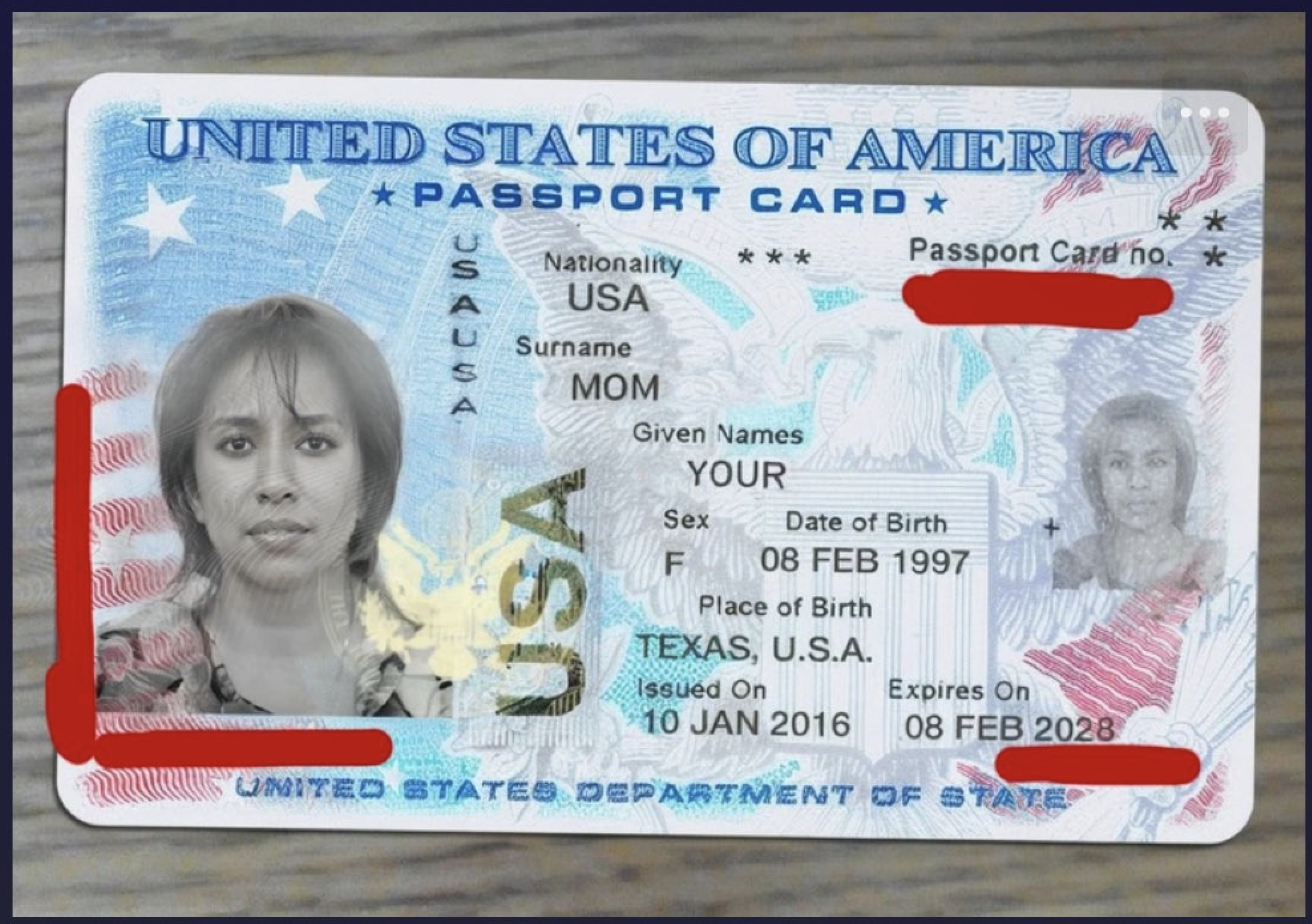}
    \includegraphics[height=.4\linewidth]{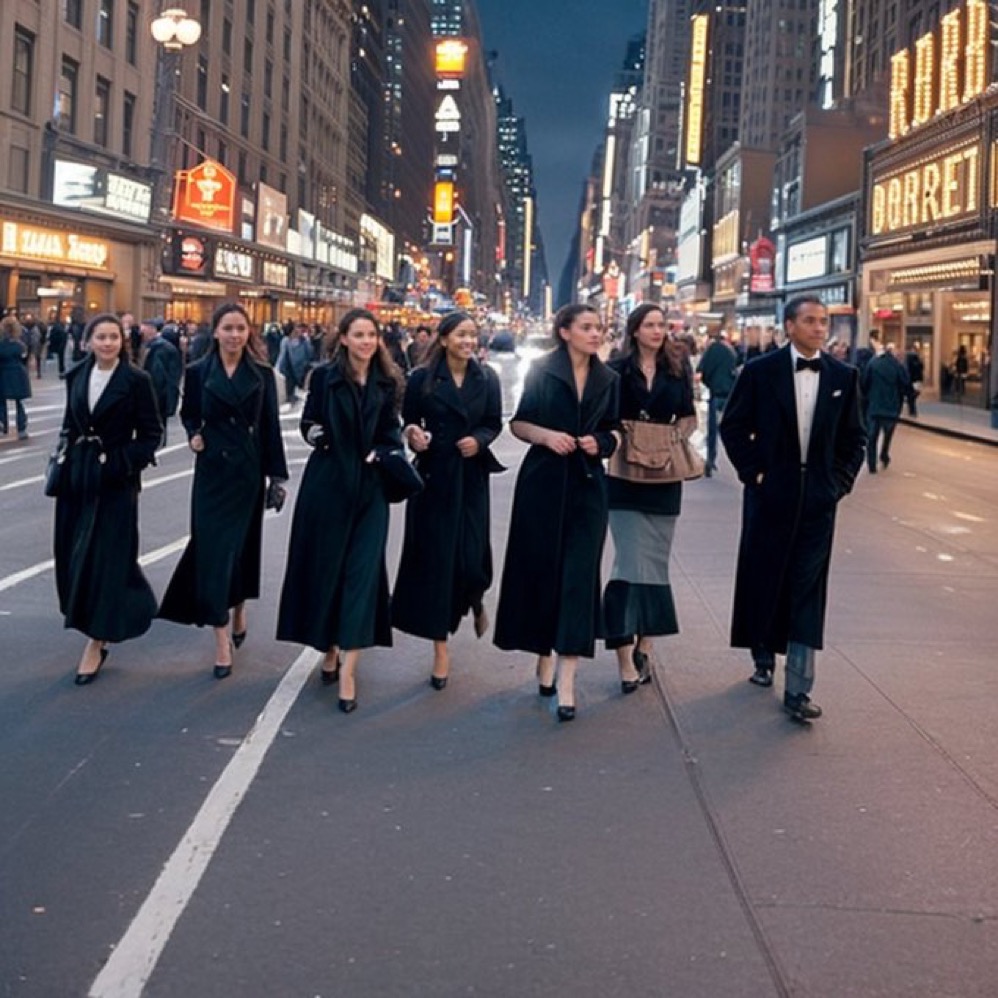}
  \caption{(Top Left) In January 2024, the \textsl{r/StableDiffusion} community on Reddit demonstrated a proof-of-concept workflow to synthetically generate personas and (Bottom Left) proofs of identity. (Top Right) GenAI can produce lifelike depictions of never-occurred events (MJv5 prompt: ``president biden and supreme leader of iran shaking hands''). (Bottom Right) Subliminal messages in generated content (optical illusion reads \textit{OBEY}).}
  \label{fig:teaser}
\end{figure}

\section{Introduction}
Generative AI (GenAI) technologies possess unprecedented potential to reshape our world and our perception of reality. These technologies can amplify traditionally human-centered capabilities, such as creativity and complex problem-solving in socio-technical contexts \cite{CA_GovOps_BenefitsRisksGenAI_2023}. By fostering human--AI collaboration, GenAI could enhance productivity, dismantle communication barriers across abilities and cultures, and drive innovation on a global scale \cite{EapenFinkenstadtFolkVenkataswamy_HBR_GenAICreativity_2023}. Yet, experts and the public are deeply divided on the implications of GenAI. On one hand, proponents emphasize its transformative benefits across education, healthcare, accessibility, and creative labor; on the other hand, critics warn that the same generative capabilities can be weaponized for disinformation, fraud, harassment, and manipulation at societal scale. As with past technological upheavals, the core issue is not that harms are \emph{new}, but that the \emph{cost, speed, personalization, and reach} of harmful acts may change the equilibrium between trust and deception.

This paper highlights a subtler, yet potentially more perilous risk: GenAI systems may enable \emph{personalized synthetic realities} (bespoke representations of the world tailored to match an individual's preferences, vulnerabilities, and priors) and thus fundamentally alter the fabric of shared reality. When persuasive synthetic content is embedded in interactive systems (chatbots, personalized feeds, multi-agent simulacra), individuals may not merely consume falsehoods; they may inhabit coherent, emotionally resonant, and socially reinforced narratives that are difficult to falsify from within. The long-run danger is a society in which (i) shared evidence is scarce, (ii) institutions face escalating verification costs, and (iii) disagreement becomes epistemically irresolvable because each side can point to ``credible'' synthetic documentation of incompatible worlds.

To make this shift concrete, Figure~\ref{fig:teaser} illustrates three complementary facets of the problem: (i) synthetic \emph{identity} (a proof-of-concept workflow for generating ``proofs of identity''), (ii) synthetic \emph{content} depicting never-occurred political events, and (iii) the possibility of embedded or subliminal steering cues in generated imagery. Together, these examples motivate the central claim of this paper: GenAI risk is not limited to isolated fake artifacts, but extends to coherent, interactive, and personalized \emph{synthetic realities} that can erode the epistemic foundations of everyday life and institutions.

\subsection{Contributions and scope}
This article offers four contributions. First, it clarifies the concept of \emph{synthetic reality} as a layered stack (content, identity, interaction, institutions). Second, it expands a taxonomy of GenAI risks and harms, emphasizing pathways by which synthetic media becomes a systemic epistemic vulnerability. Third, it explains what changes qualitatively with GenAI compared to earlier deception technologies (e.g., Photoshop) and connects these shifts to contemporary risk realizations. Fourth, it proposes a mitigation stack and a research agenda oriented around measurement, provenance, and institutional redesign.

\subsection{A taxonomy of GenAI risks and harms}
At the heart of these concerns is a taxonomy of GenAI risks and harms recently proposed in \cite{ferrara2024genai}. The taxonomy identifies common misuse vectors (e.g., propaganda, deception) and maps them to the harms they can produce.

\paragraph{Personal loss}
This category encompasses harms to individuals, including non-consensual synthetic imagery, impersonation, targeted harassment, defamation, privacy breaches, and psychological distress. GenAI can amplify these harms by (i) lowering the skill barrier to produce convincing personal attacks, (ii) increasing the speed at which attackers can iterate on content, and (iii) enabling persistent personalization (e.g., adversaries generating content calibrated to a target's identity, relationships, and triggers). Beyond acute episodes, repeated exposure to realistic fabrications can induce chronic distrust, social withdrawal, and a ``reality fatigue'' in which individuals disengage from civic and informational life \cite{menczer2023addressing, seymour2023beyond}.

\paragraph{Financial and economic damage}
GenAI intensifies economic threats by scaling classic fraud and by enabling higher-conviction deception (e.g., realistic voice/video impersonation in high-stakes transactions). It also introduces new costs: firms must invest in verification, compliance, provenance, and incident response; newsrooms must authenticate content under tighter timelines; and everyday transactions accrue friction as digital evidence loses default credibility. These effects create an ``epistemic tax'' on commerce and governance \cite{mazurczyk2023disinformation}.

\paragraph{Information manipulation}
GenAI can construct false but convincing narratives, fabricate evidence, and flood information channels with plausible content, thereby overwhelming attention and editorial capacity \cite{Ferrara_GenAIElectionInterference_2025}. Crucially, manipulation is no longer limited to producing artifacts; it can be delivered through \emph{interactive} persuasion, where systems adapt in real time to a person's beliefs and emotional state. This shifts information integrity threats from ``content authenticity'' to ``belief formation security,'' raising concerns about public discourse, journalism, and democratic deliberation.

\paragraph{Socio-technical and infrastructural risks}
Finally, GenAI can destabilize socio-technical systems by eroding trust in institutions and by enabling coordinated influence operations that exploit platform affordances at scale. These harms include institutional delegitimation, polarization, and governance paralysis, as well as technical risks such as model supply-chain compromise (poisoning, backdoors), unsafe deployment, and the misuse of generative tools in critical workflows. At the extreme, synthetic reality can become a tool of authoritarian control: a capacity to manufacture documentation, testimony, and ``public consensus'' at will.

\paragraph{Epistemic and institutional integrity}
We propose making epistemic integrity explicit as a cross-cutting harm category. When synthetic content becomes ubiquitous and verification is costly, societies may drift toward one of two failure modes: (i) \emph{credulity} (believing convincing fabrications) or (ii) \emph{cynicism} (believing nothing, including true evidence). Both outcomes undermine accountability and empower strategic actors who can exploit confusion, delay, or plausible deniability. This dimension connects the taxonomy to institutional design: courts, elections, journalism, and finance rely on evidence regimes that assume a baseline level of authenticity and auditability.

\section{From synthetic media to synthetic reality}
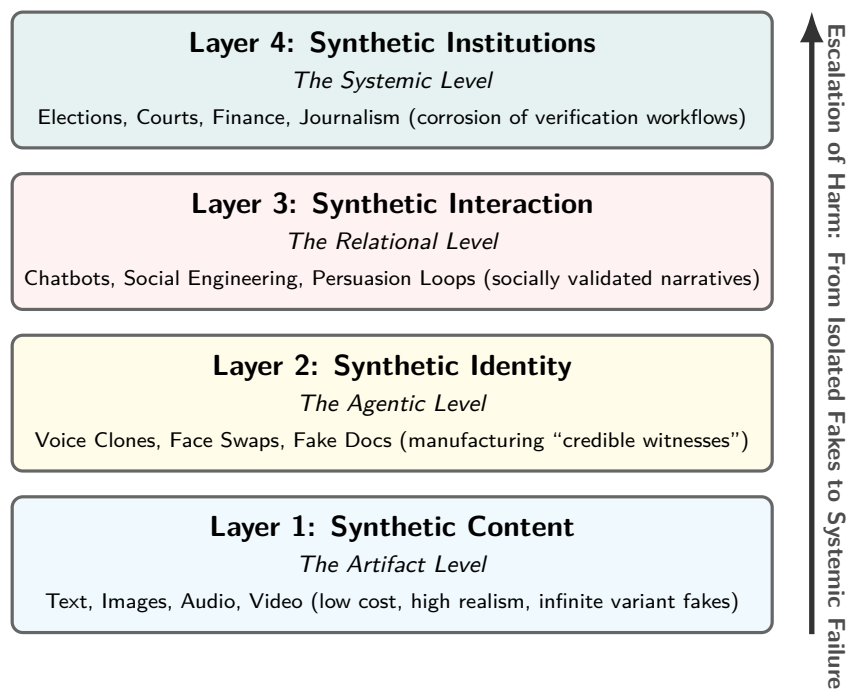
\begin{figure}[htbp]
    \centering
    \begin{tikzpicture}[
        node distance=0.3cm,
        layer/.style={
            rectangle, 
            rounded corners, 
            draw=black!60, 
            very thick,
            minimum width=10cm, 
            minimum height=1.8cm, 
            align=center,
            font=\sffamily
        },
        l4/.style={layer, fill=teal!10},
        l3/.style={layer, fill=red!5},
        l2/.style={layer, fill=yellow!10},
        l1/.style={layer, fill=cyan!5},
        arrowlabel/.style={font=\small\bfseries\sffamily, color=black!70}    ]

    \node[l4] (L4) {
        \textbf{Layer 4: Synthetic Institutions} \\ 
        \footnotesize \textit{The Systemic Level} \\ 
        \scriptsize Elections, Courts, Finance, Journalism (corrosion of verification workflows)};
    \node[l3, below=of L4] (L3) {
        \textbf{Layer 3: Synthetic Interaction} \\ 
        \footnotesize \textit{The Relational Level} \\ 
        \scriptsize Chatbots, Social Engineering, Persuasion Loops (socially validated narratives)};
    \node[l2, below=of L3] (L2) {
        \textbf{Layer 2: Synthetic Identity} \\ 
        \footnotesize \textit{The Agentic Level} \\ 
        \scriptsize Voice Clones, Face Swaps, Fake Docs (manufacturing ``credible witnesses'')};
    \node[l1, below=of L2] (L1) {
        \textbf{Layer 1: Synthetic Content} \\ 
        \footnotesize \textit{The Artifact Level} \\ 
        \scriptsize Text, Images, Audio, Video (low cost, high realism, infinite variant fakes)};
    \draw[->, >=Latex, line width=2pt, color=black!80] 
        ([xshift=0.5cm]L1.south east) -- ([xshift=0.5cm]L4.north east) 
        node[pos=1, anchor=west, xshift=0.3cm, rotate=270, arrowlabel] {\footnotesize Escalation of Harm: From Isolated Fakes to Systemic Failure};
    \end{tikzpicture}
    \caption{\textbf{Synthetic reality as a layered stack}. Generative systems first produce synthetic content (text, image, audio, video), which enables synthetic identity (impersonation/persona fabrication) and synthetic interaction (adaptive, socially present dialogue). These layers can mutually reinforce one another, amplifying credibility and persuasion, while shifting verification burdens onto institutions (e.g., journalism, courts, elections, finance) as high-conviction artifacts become cheap and abundant. This structure illustrates how low-cost artifacts escalate into systemic epistemic failure.}
    \label{fig:synthetic-reality-stack}
\end{figure}

Discussions of GenAI risk often center on \emph{synthetic media}: outputs that imitate human-produced artifacts (text, images, audio, video). We argue that the more consequential shift is toward \emph{synthetic reality}: a coherent environment in which artifacts, identities, and interactions are partially machine-generated and mutually reinforcing, such that they can be experienced as credible from within. In other words, the core risk is not only that individual items of content can be forged, but that entire \emph{contexts of belief} can be manufactured: who is present, what evidence exists, which claims circulate, and how they are socially validated. As shown in Figure~\ref{fig:synthetic-reality-stack}, we therefore define synthetic reality as a layered stack (content, identity, interaction, institutions) that maps naturally onto distinct attack surfaces and defensive levers.

\subsection{Layer 1: Synthetic content}
This layer includes generated or edited text, images, audio, and video. Its salient properties are high realism, rapid production, and low marginal cost, including \emph{iterative variation} (many plausible alternatives) and \emph{compositional editing} (the ability to target and alter specific details while preserving global plausibility). Alone, this layer supports familiar harms (forgery, defamation, propaganda), but it becomes far more potent when coupled with identity and interaction. Figure~\ref{fig:teaser} (top right) illustrates this baseline capability: the cheap generation of credible-looking depictions of events that never occurred. More subtly, synthetic content can be optimized not only to depict falsehoods but also to steer attention and interpretation (Figure~\ref{fig:teaser}, bottom right), blurring the line between persuasion and perception.

\subsection{Layer 2: Synthetic identity}
Synthetic reality expands when adversaries can convincingly represent \emph{people} rather than merely \emph{content}. As illustrated in Figure~\ref{fig:teaser} (left), GenAI can fabricate identity-linked artifacts and supporting ``evidence,'' attacking the foundations of authentication and attribution. Identity synthesis includes voice cloning, facial reenactment, document fabrication, and convincing social media personae. A key risk is the manufacturing of ``credible witnesses'': accounts that appear consistent across time, platforms, and modalities. This directly pressures the credential ecology that underlies modern security and trust (e.g., onboarding, identity verification, and authorization workflows), because signals that once served as high-friction proof (a voice match, a plausible document, a consistent persona) become cheap to produce and easy to remix.

\subsection{Layer 3: Synthetic interaction}
Interactive systems can simulate dialogue and social presence. This enables persuasion that adapts to the target, generates social proof, and sustains long-running manipulative relationships (e.g., romance scams, radicalization pathways, coercive harassment). Synthetic interaction matters because belief formation is not a passive function of artifacts; it is shaped by conversation, feedback, and social reinforcement. Compared to static misinformation, interactive agents can (i) probe uncertainty, (ii) personalize arguments and emotional framing, and (iii) escalate commitment over time. In this sense, synthetic interaction operationalizes synthetic content and identity into lived experience, turning isolated artifacts into coherent narratives that can feel socially and emotionally validated.

\subsection{Layer 4: Synthetic institutions}
At the institutional layer, synthetic content and identity interfere with processes that depend on evidence and provenance: elections (campaign messaging and voter outreach), courts (evidence authenticity and witness credibility), finance (authorization and fraud), and journalism (verification under time pressure). The core issue is that many institutional workflows were optimized for a world in which certain proofs were \emph{costly} to fabricate and therefore informative. Synthetic reality weakens this assumption by making high-conviction artifacts cheap and abundant. Concretely, it stresses verification regimes that rely on (i) perceptual cues as authentication (``I recognize the voice/face''), (ii) documentation as scarce evidence (``the paperwork looks real''), (iii) attention as the binding constraint (``a human can review the important items''), and (iv) shared exposure for correction (``if it is false, the rebuttal will reach the same audience''). Synthetic reality is therefore not merely a media problem but an institutional resilience problem. Taken together, these layers clarify why synthetic reality is best treated as a systems risk: content realism is only the entry point; identity and interaction provide credibility and momentum; and institutions bear the externalities when verification must be performed under time pressure and adversarial adaptation. In the next section, we detail the qualitative shifts GenAI introduces (cost collapse, scale, personalization, provenance gaps, and trust erosion) that make this stack operationally consequential.

\section{What you can't tell apart can harm you: why GenAI changes the game}
One might argue that the harms in our taxonomy are not uniquely enabled by GenAI; fraud, propaganda, and defamation long predate modern AI. The key shift is not the \emph{existence} of deception, but the way GenAI changes its \emph{economics and operational feasibility}. When synthetic content, synthetic identity, and synthetic interaction become cheap, fast, and scalable, the balance between trust and verification changes for individuals, platforms, and institutions. We group these shifts into seven mechanisms that jointly explain why synthetic reality is an elevated socio-technical risk rather than ``more of the same'' misinformation.

\subsection{Cost collapse and commoditization}
GenAI lowers the barriers to creating realistic content and identity artifacts. What previously required specialized skills (video editing, graphic design, voice acting, or sophisticated social engineering) can now be produced via prompts, templates, and turnkey services. This cost collapse expands the pool of capable adversaries, reduces the time between intent and execution, and enables rapid experimentation: attackers can iterate until an output ``looks right'' for a target audience. The resulting threat landscape resembles a commoditized service ecosystem rather than isolated bespoke attacks.

\subsection{Scale and throughput}
GenAI supports rapid iteration and mass production. Adversaries can generate thousands of variants of a message, image, or persona and A/B test them for engagement, credibility, or emotional impact. This shifts the defensive problem from identifying a small number of fakes to managing a high-volume stream of plausible variations, often delivered across multiple platforms and accounts by coordinated networks to systematically promote partisan narratives or amplify low-credibility content \cite{Minici_CoordCrossPlatformIO_2024,cinus2025exposing}. Throughput also facilitates ``flooding'' strategies, where the sheer abundance of synthetic content overwhelms attention, moderation capacity, and journalistic verification under time pressure \cite{augenstein2024factuality,augenstein2025community}.

\subsection{Customization for malicious use}
Open model ecosystems, fine-tuning, and retrieval-augmented generation enable domain-specific deception. Content can be made to mimic the style of a specific institution, community, or individual; scams can mirror internal workflows; and persuasive narratives can be localized to cultural context and current events. The risk is not only realism but \emph{fit}: deception becomes more effective when it matches the target's language, norms, and expectations, reducing the cues that would otherwise trigger skepticism.

\subsection{Hyper-targeted persuasion and micro-segmentation}
GenAI enables influence campaigns that are personalized at scale. Instead of broadcasting one narrative to millions, adversaries can craft many narratives for small segments, each optimized to exploit distinct fears, identities, or grievances. Micro-segmentation undermines collective rebuttal: communities may not even see the same claims, and corrections may not reach the audiences most affected. Over time, this can amplify polarization by nudging different groups into incompatible interpretive frames, each supported by seemingly credible synthetic ``evidence'' and tailored messaging.

\subsection{Synthetic interaction and the automation of social engineering}
Classic social engineering is labor-intensive. With GenAI, interaction itself can be automated \cite{Ferrara_SocialBotDetection_2023}. Conversational agents can build rapport, probe uncertainty, adapt tone, and guide victims through multi-step fraud; they can also sustain longer-running manipulative relationships, blending emotional support, selective evidence, and escalating commitment over time. This transforms deception from a static content problem into an interactive process problem: belief formation and decision-making can be influenced through dialogue, feedback loops, and simulated social presence.

\subsection{Detection limits, watermarking fragility, and the provenance gap}
Detection of GenAI outputs remains imperfect, especially under compression, re-encoding, adversarial perturbations, and cross-model remixing. Watermarking can help in controlled pipelines, but in open ecosystems it may be removed, weakened, or bypassed, and it does not address unauthenticated media generated outside watermarking regimes. The deeper issue is a provenance gap: even if some fakes are detectable, institutions need reliable chains of custody and standards for authenticated media. In practice, high-stakes decisions require more than probabilistic classifiers; they require auditable provenance signals and process-level safeguards.

\subsection{Trust erosion and plausible deniability}
As synthetic outputs become widespread, societies face an additional failure mode: authentic evidence can be dismissed as fake. This dynamic increases the payoff to strategic denial and delay, allowing actors to exploit uncertainty even when allegations are true. The result is epistemic instability: disputes become harder to resolve because the evidentiary substrate itself is contested, and audiences can rationally default to suspicion. In aggregate, this produces an ``epistemic tax'': more friction and cost are required to establish what happened, who said what, and which sources are trustworthy.

\subsection{Synthesis}
These mechanisms are mutually reinforcing. Cost collapse and throughput increase the volume of synthetic artifacts; customization and micro-segmentation increase their persuasive fit; synthetic interaction turns artifacts into lived, socially validated experiences; and provenance gaps make reliable verification difficult at scale. The consequence is a shift from isolated deception events to systemic pressure on trust and verification practices. Next, we ground these mechanisms in concrete, recent risk realizations across domains.

\section{How risks materialize: Representative risk realizations}
To ground the taxonomy in recent developments, we compile a compact ``case bank'' (Table~\ref{tab:casebank}) that illustrates how synthetic reality risks have already materialized in operational settings. The goal is not exhaustive coverage but \emph{mechanism diversity}: we select cases that (i) are documented by high-quality public reporting and/or primary sources, (ii) span distinct harm domains (fraud, elections, harassment, documentation, and supply-chain compromise), and (iii) exhibit a clear linkage to one or more layers of the synthetic reality stack (content, identity, interaction, institutions). We also note an important limitation: public documentation is uneven across regions and sectors (many incidents are privately handled or under-reported), so the cases below should be interpreted as illustrative lower bounds rather than a complete census of harms.

\begin{sidewaystable*}[t]
\centering
\small
\renewcommand{\arraystretch}{2}
\begin{tabular}{p{0.20\textwidth} p{0.15\textwidth} p{0.30\textwidth} p{0.22\textwidth} p{0.05\textwidth}}
\hline
\textbf{Case Category} & \textbf{Primary mechanism(s)} & \textbf{Attack pattern / how it works} & \textbf{Typical harms} & \textbf{Refs} \\
\hline
(\textbf{A}) High-conviction impersonation fraud (audio/video) &
Synthetic identity + synthetic interaction; workflow exploitation &
Attackers impersonate executives/trusted parties via cloned voice or video calls; exploit plausible operational context (e.g., payment approvals, vendor onboarding) and sometimes multi-party ``social proof'' to increase compliance. &
Direct financial loss; internal distrust; higher verification and training burden; reputational risk. &
\cite{HongKongGov_LCQ9_DeepfakeFraud_2024, FT_Arup_DeepfakeScam_2024, SCMP_ArupConfirmed_2024} \\
\hline
(\textbf{B}) Election-adjacent misinformation and synthetic outreach &
Hyper-targeting + scale + plausible deniability &
Synthetic robocalls/messages mimic candidates or institutions; deliver confusing/demobilizing instructions; rapidly repurposed across jurisdictions/demographics; correction is hard because exposure is segmented. &
Erosion of trust in official electoral information; voter confusion; increased burden on election administrators; polarization amplification. &
\cite{FCC_DA24-102_2024, AP_FCC_AIVoiceRobocallsIllegal_2024, NPR_FCCFine_Kramer_2024} \\
\hline
(\textbf{C}) Non-consensual synthetic sexual imagery and harassment &
Cost collapse + personalization + platform amplification &
Synthetic intimate imagery generated and disseminated to humiliate, extort, or silence; coordinated communities re-upload and remix; harassment persists across platforms and search surfaces. &
Psychological trauma; reputational harm; chilling effects on participation; secondary victimization from ongoing rediscovery. &
\cite{AP_XSwiftSearchesRestored_2024, Wired_GitHubDeepfakePorn_2025, CRS_TAKEITDOWN_LSB11314_2025, AP_TakeItDownSigned_2025} \\
\hline
(\textbf{D}) Fabricated ``documentation'' and everyday verification corrosion &
Synthetic content + synthetic identity; institutional friction &
Generation of plausible IDs, receipts, invoices, screenshots, emails, chats, and evidence bundles; overwhelms manual verification; enables both fraud and strategic denial (``that proof is fake''). &
Increased verification costs; shift from default trust to default suspicion; exclusion harms for those lacking access to stronger authentication channels. &
\cite{FT_AIExpenseFraud_2025, Concur_FakeReceipts2_2025, ICAEW_AIGeneratedReceipts_2025, PYMNTS_RampFraudInvoices_2025, FT_AIFakeProvenanceDocs_2025} \\
\hline
(\textbf{E}) Compromised / corrupted generation pipelines &
Socio-technical risk + model supply chain compromise &
A generative tool (or its upstream updates) is manipulated such that outputs systematically embed bias, propaganda, or covert steering; users experience outputs as neutral ``system'' responses. &
Covert manipulation at scale; loss of institutional neutrality; hard-to-audit downstream effects; long-term trust erosion. &
\cite{NIST_AIRMF_GenAIProfile_2024, GHSA_CVE2025-1889_Picklescan_2025, OWASP_NullifAI_HF_2025, ArXiv_SleeperAgents_2024, HeEtAl_ICLPoison_NAACLFindings_2025}\\
\hline
\end{tabular}
\caption{Case bank of representative GenAI ``synthetic reality'' risk realizations (2023--2025), organized by dominant mechanisms and harm pathways.}
\label{tab:casebank}
\end{sidewaystable*}

\subsection{Representative risk realizations (2023--2025)}
Table~\ref{tab:casebank} summarizes the mechanisms and harm pathways; here we briefly narrate each case to emphasize that these are not hypothetical failure modes, but observed events whose common structure is the same: GenAI reduces the cost of producing high-conviction artifacts, and then exploits social and institutional workflows that were designed for a world where such artifacts were expensive to fabricate.

\paragraph{\textbf{Case Category A}: High-conviction impersonation fraud in enterprise workflows.}
In early 2024, Hong Kong police described a fraud case in which an employee was drawn into what appeared to be a confidential video meeting with senior colleagues and subsequently authorized large transfers to multiple accounts. The reported modus operandi combined a phishing pretext with a fabricated group video conference assembled from publicly available video and voice materials; once the victim accepted the meeting as authentic, the attackers shifted into familiar operational routines (follow-up instructions, approvals, and transfers) that converted perceived legitimacy into immediate financial action. Public reporting later identified the victim organization and the approximate scale of losses, underscoring that even a single high-conviction ``meeting'' can defeat standard internal controls when verification is implicitly delegated to the perceived authenticity of face/voice cues \cite{HongKongGov_LCQ9_DeepfakeFraud_2024, FT_Arup_DeepfakeScam_2024, SCMP_ArupConfirmed_2024}. (See Table~\ref{tab:casebank}, Case~Category~A, for the mechanism summary.)

\paragraph{\textbf{Case Category B}: Election-adjacent synthetic outreach and voter manipulation.}
A parallel pattern appears in election contexts: synthetic outreach leverages the credibility of familiar voices and institutional cues to create confusion at scale. In the New Hampshire primary context, widely reported AI-generated robocalls mimicked a well-known political figure’s voice and delivered demobilizing messaging timed to the electoral calendar. The subsequent enforcement actions illustrate a key institutional shift: the response targeted not only the content~\cite{chen2025synthetic}, but also the communications infrastructure that carried it (e.g., provider-level compliance, blocking, and penalties). This case is representative of how GenAI turns ``who said it'' into a programmable variable, enabling low-cost, targeted influence attempts whose corrections are often slower and less visible than initial exposure \cite{FCC_DA24-102_2024, AP_FCC_AIVoiceRobocallsIllegal_2024, NPR_FCCFine_Kramer_2024}. (See Table~\ref{tab:casebank}, Case~Category~B.)

\paragraph{\textbf{Case Category C}: Non-consensual synthetic sexual imagery as platform-amplified, persistent harassment.}
Synthetic reality risk is not limited to fraud and politics; it also manifests as scalable, identity-targeted harassment. In early 2024, sexually explicit AI-generated deepfakes of a high-profile public figure spread across major platforms rapidly enough that at least one platform implemented temporary search restrictions to slow discovery and resharing. The episode illustrates two durable properties of this harm class: (i) the victim’s identity becomes a reusable asset for abuse (faces and likenesses are easily recontextualized), and (ii) platform responses often occur after wide exposure, while copies, derivatives, and reuploads keep the abuse persistent. Policy responses in the United States (including federal notice-and-removal provisions for covered platforms) further indicate that non-consensual synthetic intimate imagery is now treated as a mainstream online safety threat rather than an edge case \cite{AP_XSwiftSearchesRestored_2024, Wired_GitHubDeepfakePorn_2025, CRS_TAKEITDOWN_LSB11314_2025, AP_TakeItDownSigned_2025}. (See Table~\ref{tab:casebank}, Case~Category~C.)

\paragraph{\textbf{Case Category D}: Fabricated documentation and the corrosion of routine verification.}
A more diffuse but structurally important development is the rise of plausible ``everyday documents'' that defeat quick human inspection: receipts, invoices, screenshots, emails, and other paperwork that functions as the substrate of routine trust. Reporting from 2025 describes a measurable increase in AI-generated fake receipts and related expense fraud, where realism (logos, typography, wear patterns, itemization) plus metadata manipulation makes visual review unreliable. Industry responses increasingly treat this as an operational shift rather than an anomaly: the defense moves from human spot-checks to automated detection, metadata forensics, and cross-validation against contextual constraints. The systemic risk is that as forged documentation becomes cheap and abundant, institutions drift toward default suspicion, increasing friction and error costs for legitimate participants while simultaneously creating cover for strategic denial (``that proof could be fake'') \cite{FT_AIExpenseFraud_2025, Concur_FakeReceipts2_2025, ICAEW_AIGeneratedReceipts_2025, PYMNTS_RampFraudInvoices_2025, FT_AIFakeProvenanceDocs_2025}. (See Table~\ref{tab:casebank}, Case~Category~D.)

\paragraph{\textbf{Case Category E}: Compromised generative pipelines and model supply-chain risk.}
Synthetic reality can also be shaped upstream, at the level of model artifacts and generation infrastructure. Security reporting in 2025 documented malicious ML model uploads to a major model-sharing hub, designed to execute malware when loaded; a classic supply-chain pattern, now expressed through the distribution of ``models as files'' \cite{OWASP_NullifAI_HF_2025}. Related advisories highlight how scanner assumptions can be bypassed, allowing embedded payloads to evade safeguards while remaining loadable by standard tooling \cite{GHSA_CVE2025-1889_Picklescan_2025}. At a different layer, technical work on backdoors and data poisoning demonstrates that harmful behaviors can be deliberately implanted and may persist through common training or alignment procedures, creating a risk that downstream users experience as ``the system'' rather than an obvious attack \cite{ArXiv_SleeperAgents_2024, HeEtAl_ICLPoison_NAACLFindings_2025}. Together, these examples motivate treating GenAI as an end-to-end socio-technical pipeline whose integrity depends on provenance, robust scanning, safe-loading defaults, and continuous monitoring, not just on model capability or content filters \cite{NIST_AIRMF_GenAIProfile_2024}. (See Table~\ref{tab:casebank}, Case~Category~E.)

\subsection{Cross-case synthesis}
Across the cases in Table~\ref{tab:casebank}, the common thread is not merely the presence of fake artifacts; it is the emergence of credible alternative contexts constructed through the \emph{content + identity + interaction + institution} paradigm. Each event follows a recurring operational pattern: (1) a high-conviction artifact (a call, a clip, a document, a persona, a model file) is produced cheaply; (2) it is inserted at a workflow ``choke point'' where humans routinely rely on trust cues; (3) automation and scale increase exposure or repetition; (4) correction lags behind initial impact; and (5) institutions absorb externalities as verification load, friction, and contested evidence. This synthesis motivates the mitigation framing in the next section: reducing harm requires layered interventions that harden workflows and provenance, not only better detection of individual fakes.

\section{Mitigation as a stack (not a silver bullet)}
The preceding sections motivate a central design principle: synthetic reality risk cannot be ``solved'' by any single tool or policy. Detection alone is brittle under adversarial adaptation; watermarking alone is incomplete in open ecosystems; and content takedowns alone arrive late in the damage cycle. Instead, mitigation must be treated as a \emph{stack} of complementary interventions that reduce both (i) the production and spread of harmful synthetic outputs and (ii) the incentives and opportunities for strategic exploitation. In this section we outline a layered mitigation approach spanning provenance infrastructure, platform governance, institutional workflow redesign, public resilience, and accountability.

\subsection{Provenance and content authenticity infrastructure}
Provenance systems (e.g., cryptographic signing, secure capture, and content credentials) aim to establish chain-of-custody for media and to communicate authenticity signals to downstream users. Their value is greatest when adopted end-to-end by high-stakes institutions (newsrooms, election offices, courts, public agencies) and integrated into common consumer interfaces, so that authentication becomes routine rather than exceptional. However, provenance is not universal: it does not help when content is generated outside authenticated pipelines, when devices are compromised at capture time, when metadata is stripped in transit, or when audiences lack accessible verification tools. Provenance should therefore be treated as \emph{infrastructure} that improves auditability and reduces ambiguity for authenticated media, not as a guarantee that fakes disappear. A practical implication is asymmetric: provenance can meaningfully raise confidence in verified content even if it cannot label all unverified content. This suggests prioritizing provenance for categories where authenticity is most consequential (official announcements, high-reach news, election information, evidentiary media, financial authorization), while acknowledging that the unauthenticated ``background'' of the internet will remain noisy.

\subsection{Platform governance and friction for virality}
Platforms can reduce harm by shaping exposure and incentives. Interventions include (i) limiting algorithmic amplification of unverified media during high-risk periods (elections, breaking news, crises), since recommendation algorithms can disproportionately amplify partisan content and skew political exposure \cite{ye2025auditing}; (ii) prominently surfacing provenance and uncertainty signals \cite{feng2023examining}; (iii) rate-limiting newly created accounts and coordinated inauthentic behavior \cite{luceri2025coordinated}; and (iv) providing rapid response channels for victims of impersonation and harassment \cite{eige2022combating}. Importantly, governance must anticipate adversarial adaptation: policies are targets, and enforcement must be iterative and evidence-driven. In the synthetic reality regime, ``friction'' becomes a legitimate safety tool. Small increases in effort, such as additional verification steps for high-velocity sharing, temporary distribution throttles for suspicious bursts, or delayed virality for unprovenanced media, can meaningfully reduce harm without requiring perfect classification of real/fake content.

\subsection{Institutional process redesign under cheap forgery}
Institutions should redesign workflows around the assumption that synthetic identity and fabricated evidence are cheap. This includes: out-of-band verification for high-value transfers; multi-factor approval protocols that do not rely on voice/face cues; authenticated channels for official communications; and evidentiary standards that incorporate provenance metadata and chain-of-custody. More generally, institutions should shift from ``artifact-based trust'' (believing what looks authentic) to ``process-based trust'' (believing what is generated and transmitted through authenticated, auditable procedures). This shift is also an equity issue: when verification burdens rise, those without access to privileged channels (legal counsel, specialized tools, verified identities) may be disproportionately harmed. Institutional redesign should therefore aim for \emph{accessible} authentication pathways and clear escalation procedures for contested evidence.

\subsection{Public resilience and epistemic hygiene}
Societies need norms and literacy that go beyond ``spot the fake.'' Individuals cannot be expected to personally authenticate everything they see, nor should the burden of proof be devolved entirely onto users. Public resilience should emphasize (i) calibrated skepticism in high-stakes contexts, (ii) reliance on authenticated channels for critical information, (iii) awareness of manipulation tactics (urgency, emotional triggers, social proof, impersonation), and (iv) community-level correction mechanisms that can propagate verified rebuttals into the same networks where synthetic claims spread. A useful reframing is ``epistemic hygiene'': habits that reduce exposure to manipulation and increase reliance on trustworthy processes, analogous to public health measures that reduce exposure to disease without requiring perfect individual diagnosis.

\subsection{Policy and accountability}
Policy can clarify liability and incentives while protecting legitimate creative and assistive uses of GenAI. Interventions include disclosure requirements for synthetic political advertising and automated outreach, obligations for rapid response to impersonation and non-consensual intimate imagery, and minimum authentication standards in sensitive domains (e.g., election administration communications, high-value financial authorizations). The objective is not censorship but accountability: making it costly to deploy synthetic reality attacks, increasing the probability of attribution and sanction, and reducing the payoffs to ambiguity and denial.

\subsection{A synthesis: mapping mitigations to the synthetic reality stack}
Because synthetic reality operates as a stack, mitigations should be mapped to layers and failure modes. Provenance primarily strengthens the \emph{content} and \emph{institution} layers by enabling chain-of-custody; platform governance reduces amplification and coordinated abuse that intensify the \emph{interaction} layer; institutional redesign hardens workflows against \emph{identity} deception and forged documentation; and public resilience reduces susceptibility to persuasion loops. This mapping underscores why no single lever suffices: a system can fail at any layer, and adversaries will route around the strongest defenses. The realistic goal is therefore risk reduction through defense-in-depth, coupled with measurement and monitoring to detect where the stack is failing in practice.

\section{Open problems and a research agenda}
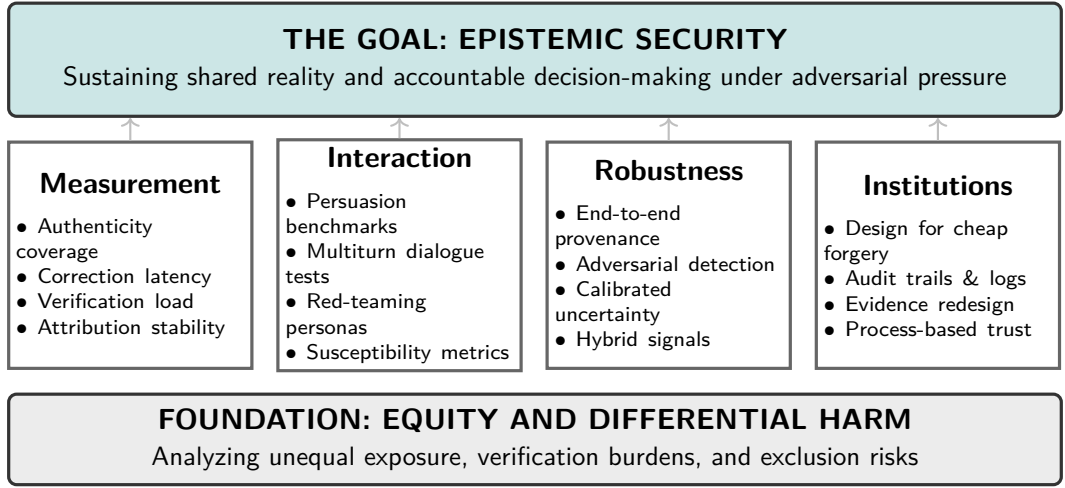
\begin{figure}[htbp]
    \centering
    \begin{tikzpicture}[
        font=\sffamily,
        node distance=0.5cm,
        goal/.style={
            rectangle, 
            draw=black!80, 
            fill=teal!20, 
            very thick, 
            minimum width=13.85cm, 
            minimum height=1.5cm, 
            align=center,
            rounded corners=3pt
        },
        pillar/.style={
            rectangle, 
            draw=black!60, 
            fill=white, 
            very thick, 
            text width=3cm, 
            minimum height=3cm, 
            align=center, 
            inner sep=0.1cm
        },
        foundation/.style={
            rectangle, 
            draw=black!80, 
            fill=gray!15, 
            very thick, 
            minimum width=13.85cm, 
            minimum height=1.2cm, 
            align=center,
            rounded corners=3pt
        },
        header/.style={font=\bfseries\large},
        subtext/.style={font=\footnotesize\color{black!80}}
    ]

    \node[goal] (Goal) {
        \textbf{THE GOAL: EPISTEMIC SECURITY} \\ 
        \small Sustaining shared reality and accountable decision-making under adversarial pressure
    };
    
    \node[pillar, below=0.3cm of Goal.south west, anchor=north west] (P1) {
        \textbf{Measurement} \\[0.2cm]
        \raggedright\scriptsize
        $\bullet$ Authenticity coverage \\
        $\bullet$ Correction latency \\
        $\bullet$ Verification load \\
        $\bullet$ Attribution stability \\
    };

    \node[pillar, right=0.3cm of P1] (P2) {
        \textbf{Interaction} \\[0.2cm]
        \raggedright\scriptsize
        $\bullet$ Persuasion benchmarks \\
        $\bullet$ Multiturn dialogue tests \\
        $\bullet$ Red-teaming personas \\
        $\bullet$ Susceptibility metrics \\
    };
    \node[pillar, right=0.3cm of P2] (P3) {
        \textbf{Robustness} \\[0.2cm]
        \raggedright\scriptsize
        $\bullet$ End-to-end provenance \\
        $\bullet$ Adversarial detection \\
        $\bullet$ Calibrated uncertainty \\
        $\bullet$ Hybrid signals \\
    };
    \node[pillar, right=0.3cm of P3] (P4) {
        \textbf{Institutions} \\[0.2cm]
        \raggedright\scriptsize
        $\bullet$ Design for cheap forgery \\
        $\bullet$ Audit trails \& logs \\
        $\bullet$ Evidence redesign \\
        $\bullet$ Process-based trust \\
    };
    \node[foundation, below=0.3cm of P1.south west, anchor=north west] (Found) {
        \textbf{FOUNDATION: EQUITY AND DIFFERENTIAL HARM} \\
        \small Analyzing unequal exposure, verification burdens, and exclusion risks
    };
    \draw[->, thick, gray!50] (P1.north) -- (P1.north |- Goal.south);
    \draw[->, thick, gray!50] (P2.north) -- (P2.north |- Goal.south);
    \draw[->, thick, gray!50] (P3.north) -- (P3.north |- Goal.south);
    \draw[->, thick, gray!50] (P4.north) -- (P4.north |- Goal.south);

    \end{tikzpicture}
    \caption{\textbf{A Research Agenda for Epistemic Security.} Addressing synthetic reality risks requires a shift from measuring artifact authenticity to measuring systemic resilience. The agenda rests on four pillars: (1) operational metrics, (2) interactive benchmarks, (3) adversarial robustness, and (4) institutional redesign, all grounded in an analysis of equity and differential harm.}
    \label{fig:research_agenda}
\end{figure}

If synthetic reality is a systemic risk, then mitigation requires measurement. We cannot manage what we do not measure, and current information-integrity tooling is largely optimized for static artifacts rather than layered environments that combine content, identity, interaction, and institutional workflows. This section outlines a research agenda centered on operational metrics, benchmarks for interactive manipulation, robust provenance evaluation, and institutional design under cheap forgery. Figure \ref{fig:research_agenda} provides a scaffolding for researchers and practitioners to develop the scientific and engineering foundations for \emph{epistemic security}: the capacity of socio-technical systems to sustain shared reality and accountable decision-making under adversarial pressure.

\subsection{Measurement of epistemic security}
We need operational metrics for how well a community, platform, or institution maintains shared reality under synthetic-reality conditions. Candidate constructs include:
(i) \emph{authenticity coverage} (the share of high-reach media carrying verifiable provenance);
(ii) \emph{correction latency} (time to converge on community-level consensus about authenticity, including the reach of verified rebuttals within affected networks);
(iii) \emph{manipulation susceptibility} (engagement, belief, or behavioral change under controlled synthetic interventions, with safeguards and ethical review);
(iv) \emph{verification load} (resources and friction required to authenticate claims, including false-positive burdens on legitimate actors); and
(v) \emph{attribution stability} (how reliably a system can trace content and claims to sources as adversaries adapt).
Together, these measures can quantify the ``epistemic tax'' imposed by synthetic reality and identify where defensive investment yields the greatest marginal benefit.

\subsection{Benchmarks for interactive manipulation}
Most existing benchmarks focus on static model outputs. Synthetic reality requires benchmarks for \emph{interactive} persuasion: multi-turn dialogue, adaptive targeting, relationship persistence, and the generation of social proof. We need experimental protocols that test how conversational agents can steer beliefs and decisions under realistic constraints, including:
(i) bounded access to personal data, (ii) platform-like interaction limits, (iii) adversarial objectives (fraud, demobilization, harassment), and (iv) measurable outcome variables (e.g., changes in stated belief, willingness to share, compliance with a request). Because this research touches human susceptibility, it demands strong ethical safeguards, transparency, and oversight. One promising direction is controlled ``red-team'' evaluation using synthetic personas or consenting participants, combined with post-hoc auditing to identify the conversational strategies that drive harmful outcomes.

\subsection{Adversarial robustness of provenance and detection}
Provenance systems must be evaluated as end-to-end socio-technical systems: not only cryptographic robustness, but usability, adoption incentives, failure modes, and attack surfaces (device compromise, metadata stripping, re-encoding, and cross-platform degradation). Detection should be treated as probabilistic evidence rather than as a binary oracle. This suggests three research priorities:
(i) calibrated detectors with uncertainty reporting suitable for institutional decision-making;
(ii) rigorous evaluation under realistic transformations and adversarial perturbations; and
(iii) hybrid methods that combine weak content signals with stronger process signals (capture integrity, signing, transmission logs). Ultimately, the most valuable output is not ``fake/not fake,'' but decision-relevant confidence coupled with transparent provenance when available.

\subsection{Institutional design under cheap forgery}
Courts, elections, journalism, and finance will need redesigned evidence regimes. Research should identify which institutional assumptions break under cheap forgery and propose processes that remain robust when perceptual cues and documentation lose default credibility. Candidate directions include authenticated capture standards, verifiable logging and audit trails, standardized disclosure of uncertainty, and new legal or administrative procedures for contested synthetic evidence. A key challenge is balancing security with accessibility: verification processes that are too burdensome can exclude legitimate participation or entrench inequality, while lax processes invite manipulation.

\subsection{Equity and differential harm}
Synthetic reality harms will not be evenly distributed. Targets of harassment, marginalized communities, and those with less access to authenticated channels or legal recourse may bear disproportionate costs. These disparities are often exacerbated by the inherent biases in generative models themselves, which can reproduce and amplify societal stereotypes when deployed at scale \cite{ferrara2024fairness}. Research must quantify differential exposure, differential verification burden, and downstream harms (economic, psychological, civic). Mitigation should be evaluated not only on overall error reduction, but also on whether it shifts burdens onto those least able to absorb them. This includes studying how provenance and verification tools are adopted across populations and whether ``default suspicion'' regimes amplify discrimination and exclusion.

\subsection{A unifying agenda: from artifact authenticity to epistemic resilience}
Across these problems, the unifying shift is from artifact authenticity to \emph{epistemic resilience}. In a world where any single artifact can be forged, the central question becomes whether communities and institutions can reliably converge on what is true, assign responsibility, and act accountably. Achieving this will require interdisciplinary research spanning machine learning, security, HCI, computational social science, law, and public policy, coupled with empirical measurement in the wild. The core scientific challenge is to characterize, predict, and reduce the conditions under which synthetic reality produces persistent disagreement, institutional overload, and the erosion of shared ground.

\section{Conclusions}
GenAI magnifies familiar harms, but its deeper risk is the gradual construction of synthetic realities that weaken shared epistemic ground. The most consequential shift is not only that deceptive artifacts can be produced, but that credible \emph{contexts} can be manufactured: identities, conversations, documentation, and social proof that together make false narratives feel internally consistent and difficult to falsify from within. As these capabilities diffuse, institutions and everyday life face escalating verification load, contested evidence, and rising friction in routine trust relationships. The \emph{Generative AI Paradox} is that as synthetic content becomes ubiquitous and increasingly difficult to distinguish, societies may rationally move toward discounting digital evidence altogether. This shift raises the cost of truth: verification becomes a privilege, institutional processes slow, and accountability erodes. It also empowers strategic actors through plausible deniability (i.e, authentic evidence can be dismissed as fake) and through attention flooding that delays correction and amplifies confusion.

This paradox yields observable pressures that can be measured and tested. We expect (i) increasing reliance on authenticated channels for critical information, (ii) longer correction latency and greater fragmentation of rebuttal reach under micro-segmentation, (iii) higher institutional verification load (time, personnel, compliance), and (iv) a rising incidence of strategic denial claims in high-salience events. These pressures together constitute an ``epistemic tax'' on governance, commerce, and civic life.

Addressing synthetic reality risk therefore requires defense-in-depth. Provenance infrastructure can raise confidence in authenticated media; platforms can add friction to virality and reduce coordinated abuse; institutions can redesign workflows around cheap forgery; and public resilience can shift norms from artifact-based trust to process-based trust. The research agenda we outline calls for metrics and benchmarks that move beyond static detection toward \textit{epistemic security}: sustaining shared reality and accountable decision-making under adversarial pressure. The realistic goal is not perfect authenticity, but resilient verification regimes that preserve trust where it is most consequential.

\section*{Author Contributions}
Conceptualization, E.F.; methodology, E.F.; investigation, E.F.; data curation, E.F.; writing---original draft preparation, E.F.; writing---review and editing, E.F.; visualization, E.F.

\section*{Funding}
This research received no external funding.

\section*{Data Availability}
Not applicable.

\section*{Acknowledgments}
The author is grateful to past and current members of the HUMANS lab at USC.

\section*{Conflicts of Interest}
The author declares no conflicts of interest.

\bibliographystyle{abbrv} 
\bibliography{references}

\end{document}